\documentclass[iop]{emulateapj}

\usepackage{CJK}
\usepackage{txfonts}
\usepackage{graphicx}

\usepackage{graphicx}

\usepackage{soul}

\usepackage{natbib}
\usepackage{natbib}

\begin{document}
\title{Mapping the magnetic field of flare coronal loops}

\email{dak21@aber.ac.uk}

\author{D. Kuridze$^{1,2,8}$, M. Mathioudakis$^2$, H. Morgan$^1$, R. Oliver$^{3,11}$,  L. Kleint$^4,10$,  T. V. Zaqarashvili$^{5,8,9}$, A. Reid$^2$,  J. Koza$^6$, M. G. L\"ofdahl$^7$, T. Hillberg$^7$, V. Kukhianidze$^8$, and  A. Hanslmeier$^9$}
\affil{$^1$Institute of Mathematics, Physics and Computer Science, Aberystwyth University, Ceredigion, Cymru, SY23 3BZ, UK}
\affil{$^2$Astrophysics Research Centre, School of Mathematics and Physics, Queen's University Belfast, Belfast BT7 1NN, UK}
\affil{$^3$Departament de F\'{\i}sica, Universitat de les Illes Balears, E-07122 Palma de Mallorca, Spain}
\affil{$^4$University of Applied Sciences and Arts Northwestern Switzerland, Bahnhofstrasse 6, 5210 Windisch, Switzerland}
\affil{$^5$Space Research Institute, Austrian Academy of Sciences, Schmiedlstrasse 6, 8042 Graz, Austria}
\affil{$^6$Astronomical Institute, Slovak Academy of Sciences, 059 60 Tatranska Lomnica, Slovakia}
\affil{$^7$Institute for Solar Physics, Department of Astronomy, Stockholm University, Albanova University Centre, SE-106 91 Stockholm, Sweden}
\affil{$^8$Abastumani Astrophysical Observatory at Ilia State University, 3/5 Cholokashvili Avenue, 0162, Tbilisi, Georgia}
\affil{$^9$IGAM, Institute of Physics, University of Graz, Universit\"atsplatz 5, A-8010 Graz, Austria}
\affil{$^{10}$Kiepenheuer-Institut f\"ur Sonnenphysik, Sch\"oneckstrasse 6, D-79104 Freiburg, Germany}
\affil{$^{11}$Institut d'Aplicacions Computacionals de Codi Comunitari (IAC3), Universitat de les Illes Balears, Spain}


\begin{abstract}

Here we report on the unique observation of flaring coronal loops at the solar limb using high resolution imaging spectropolarimetry 
from the Swedish 1-meter Solar Telescope. The vantage position, orientation and nature of the chromospheric material that filled 
the flare loops allowed us to determine their magnetic field with unprecedented accuracy using the 
weak-field approximation method. Our analysis reveals coronal magnetic field strengths as high as 350 Gauss at 
heights up to 25 Mm above the solar limb. These measurements are substantially higher than a number of previous 
estimates and may have considerable implications for our current understanding of the extended solar atmosphere.

\end{abstract}


\section{Introduction}

The structure and dynamics of the most energetic events in the solar outer atmosphere such as flares, eruptions, coronal loops and CMEs are controlled by the magnetic field. 
This field is created deep in the solar interior by dynamo processes, transported outwards by magnetic buoyancy and 
permeates the photosphere in dense flux ropes - giving the familiar appearance of sunspots and pores, associated with coronal active regions. 
Over time, this intense emergent field diffuses throughout the photosphere and fills the solar chromosphere and corona, dominating the movement and behaviour of the hot plasma.  
The majority of the EUV/Xray emission is from loops of plasma entrained along closed fields.
Measurements of the magnetic field in loops is key to our understanding of the corona and is crucial to solve the long-standing problem of coronal heating \citep{2006SoPh..234...41K}.  
Despite the tremendous efforts made with different techniques over the past two decades, a reliable quantitative 
measurement of the magnetic flux density of coronal loops remains a central problem in solar (and stellar) physics \citep{2001STIN...0227999J,2014A&ARv..22...78W,2017SSRv..210..145C}.

\begin{figure}[b]
\begin{center}
\includegraphics[width=8.65cm]{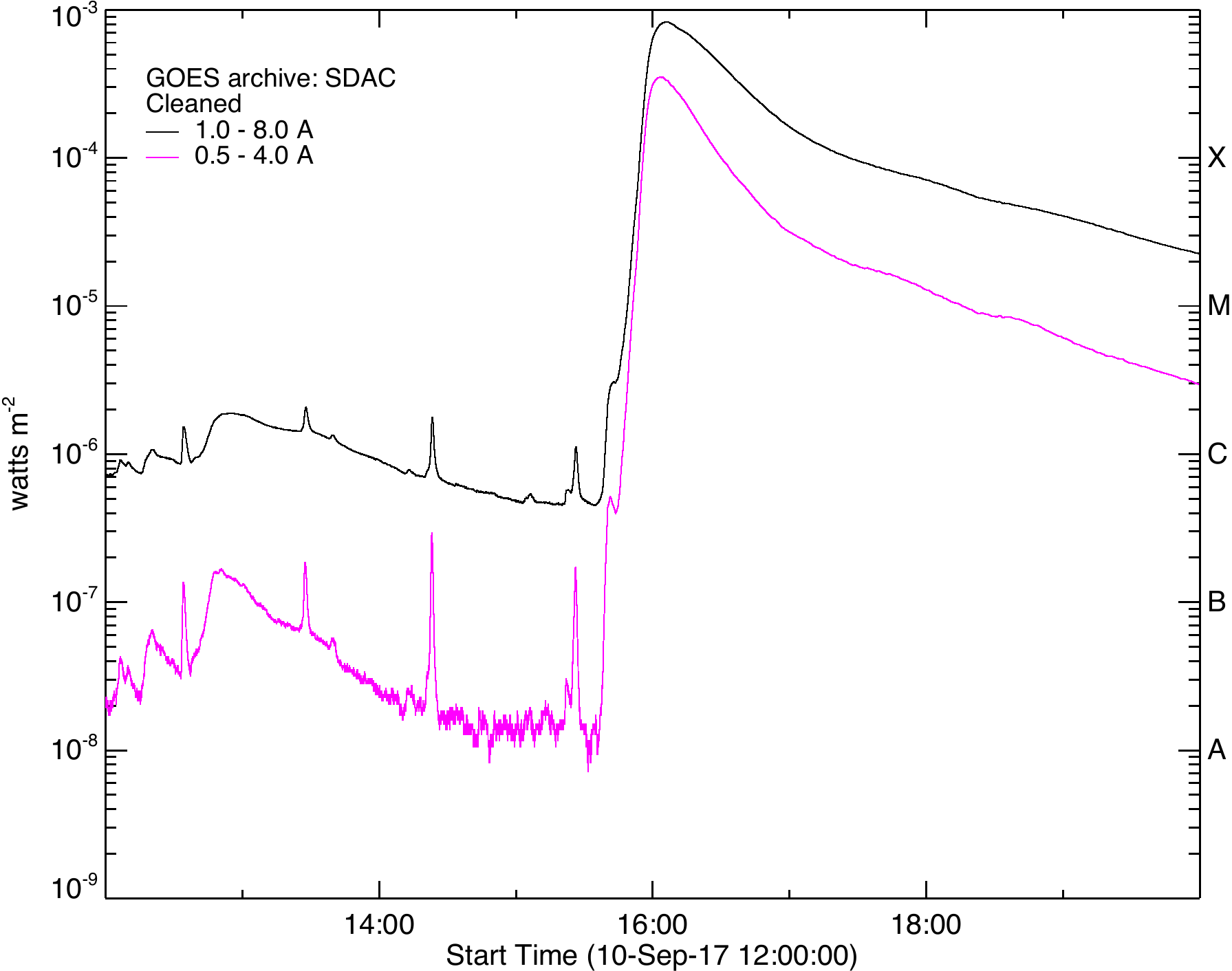}
\end{center}
\caption{GOES X-ray light curve of the X8.2 class flare of 10 September 2017.} 
\label{fig1}
\end{figure}

\begin{figure*}[t]
\begin{center}
\includegraphics[width=18.0cm]{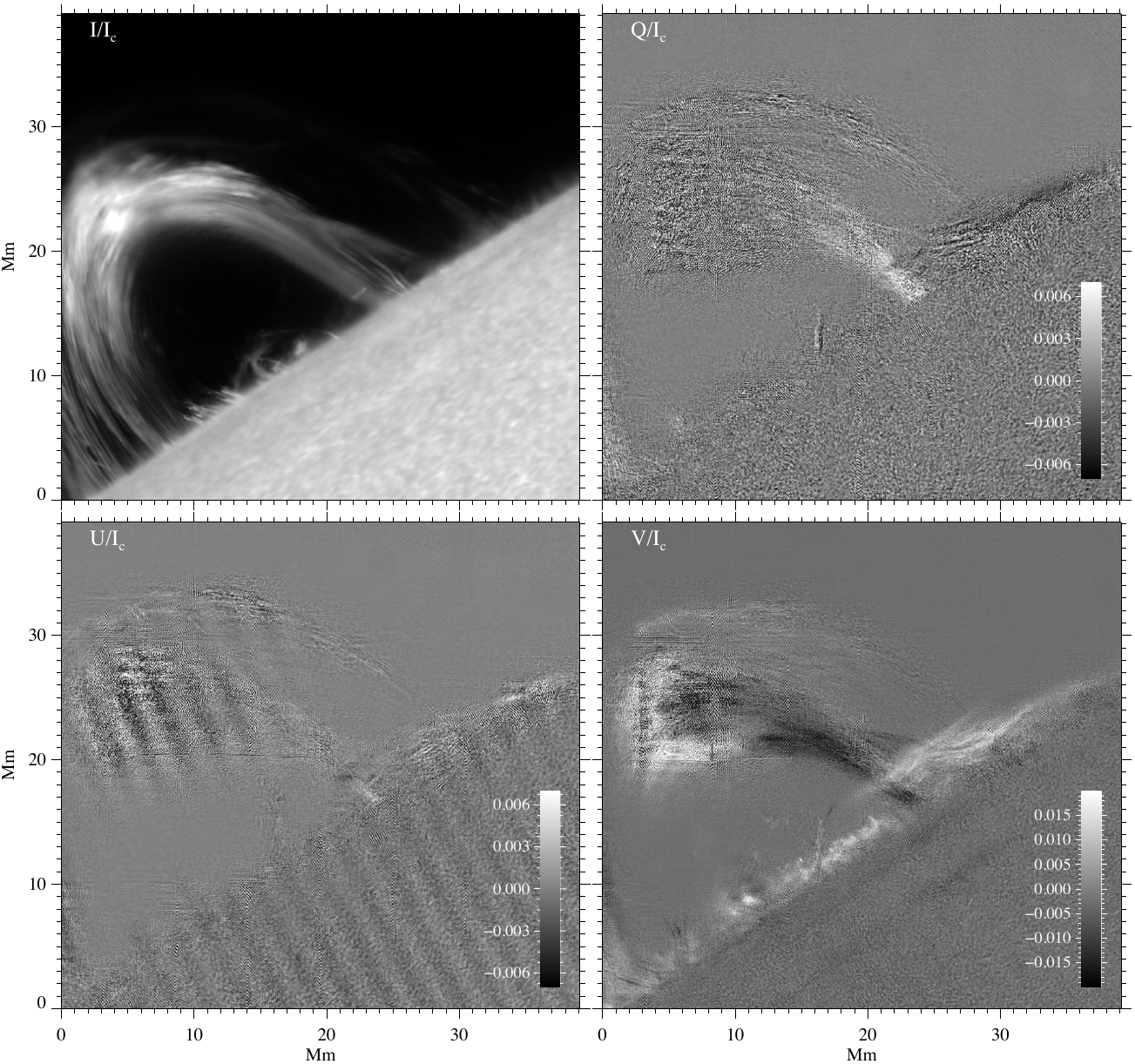}
\end{center}
\caption{SST images of the flare loops in Ca {\sc{ii}} 8542 {\AA} Stokes $I$ at $\Delta\lambda=\pm0.945~{\AA}$ and $Q, U \& V$ at $\Delta\lambda = 0.455~{\AA}$ at 16:27 UT.} 
\label{fig2}
\end{figure*}

Gyro-resonance emission is an effective method for measuring the magnetic field 
over active regions of the Sun and some stars
\citep{2006ApJ...641L..69B}. However, the radio observations are restricted by 
modest spatial resolution compared with methods in the visible and near-infrared, they can only be applied at high magnetic field strengths ($>$100 Gauss) 
to overcome free-free opacity, and their formation height is not determined by the data themselves.
Indirect methods include coronal seismology, which relies on observations of magnetohydrodynamic (MHD) waves to infer the properties of the magnetic field \citep{2005LRSP....2....3N,2007Sci...317.1192T}. 
These methods also depend on the assumed nature of the wave modes (local tube modes vs genuine Alfv\'en waves in more homogeneous media).  
The most powerful magnetic diagnostics of solar and astrophysical plasmas are based on 
measurements of the polarised states of light described by the Stokes profiles  $(I, Q, U, V)$. Despite its promise \citep{2001STIN...0227999J,2000ApJ...541L..83L,2004ApJ...613L.177L}, solar coronal polarimetry is extremely difficult due to the inherent low signal-to-noise (S/N) ratios (the corona is, at best, a million times dimmer than the photosphere). To improve the signal, a coarse spatial (5000 - 10000 km) and temporal (30 - 60 min) resolution 
has been used with small instruments in the past
\citep{2000ApJ...541L..83L,2004ApJ...613L.177L}. However, highly inhomogeneous  corona 
requires high spatial resolution 
and the highly dynamic nature of many coronal processes, such as reconnection/diffusion, requires high temporal resolution.
Current instrumentation can only achieve high resolution polarimetric measurements during certain favourable conditions. During solar flares, material evaporates from the chromosphere to the corona, and subsequently condenses and falls back to the surface along loops in the form of coronal rain. 
The high density and lower temperature (due to a lack of sustained heating) permits the use of chromospheric diagnostic transitions as the plasma falls and traces out otherwise "coronal" field lines
\citep{2016ApJ...833....5S}.

In September 2017, Active Region (AR) 12673 produced the most powerful flares of solar cycle 24 as it was rotating from disk centre to the limb. On September 10 2017, the AR was just behind the west limb when it produced an X8.2-class flare (SOL2017-09-10T16:06 UT) (Figure 1).  
Here we report on the unique observations of the flaring coronal loops during this event at the solar limb using high resolution imaging spectropolarimetry 
in the line of singly ionised calcium at 8542.1 \AA,
from the Swedish 1-meter Solar Telescope. The vantage position, orientation and nature of the chromospheric material that filled the 
flare loops allowed us to determine their magnetic field with unprecedented accuracy using weak-field approximation (WFA) method.

\section{Observations and Data Reduction}

\begin{figure*}[t]
\begin{center}
\includegraphics[width=18.1cm]{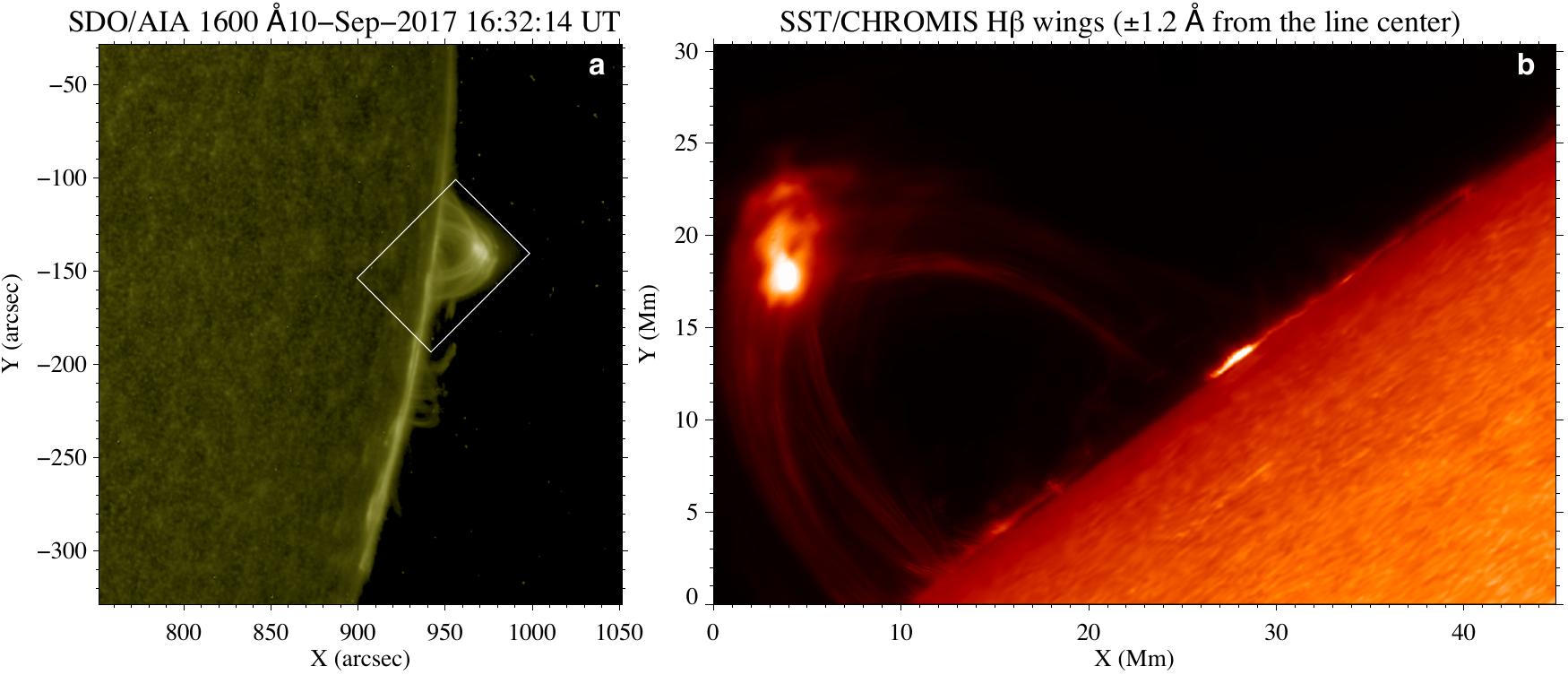}
\end{center}
\caption{SDO/AIA 1600 {\AA} image (a) of AR12673 near the west limb showing the X8.2 flare coronal loops on 10 September 2017. SST 
FOV is outlined with the white box. A composite of SST/CHROMIS H$\beta\pm1.2~{\AA}$ (coresponding to $\mathrm{\pm75~km~s^{-1}}$) images (b) of the same flaring loops.}
\label{fig1}
\end{figure*}

\begin{figure*}[t]
\begin{center}
\includegraphics[width=17.3cm]{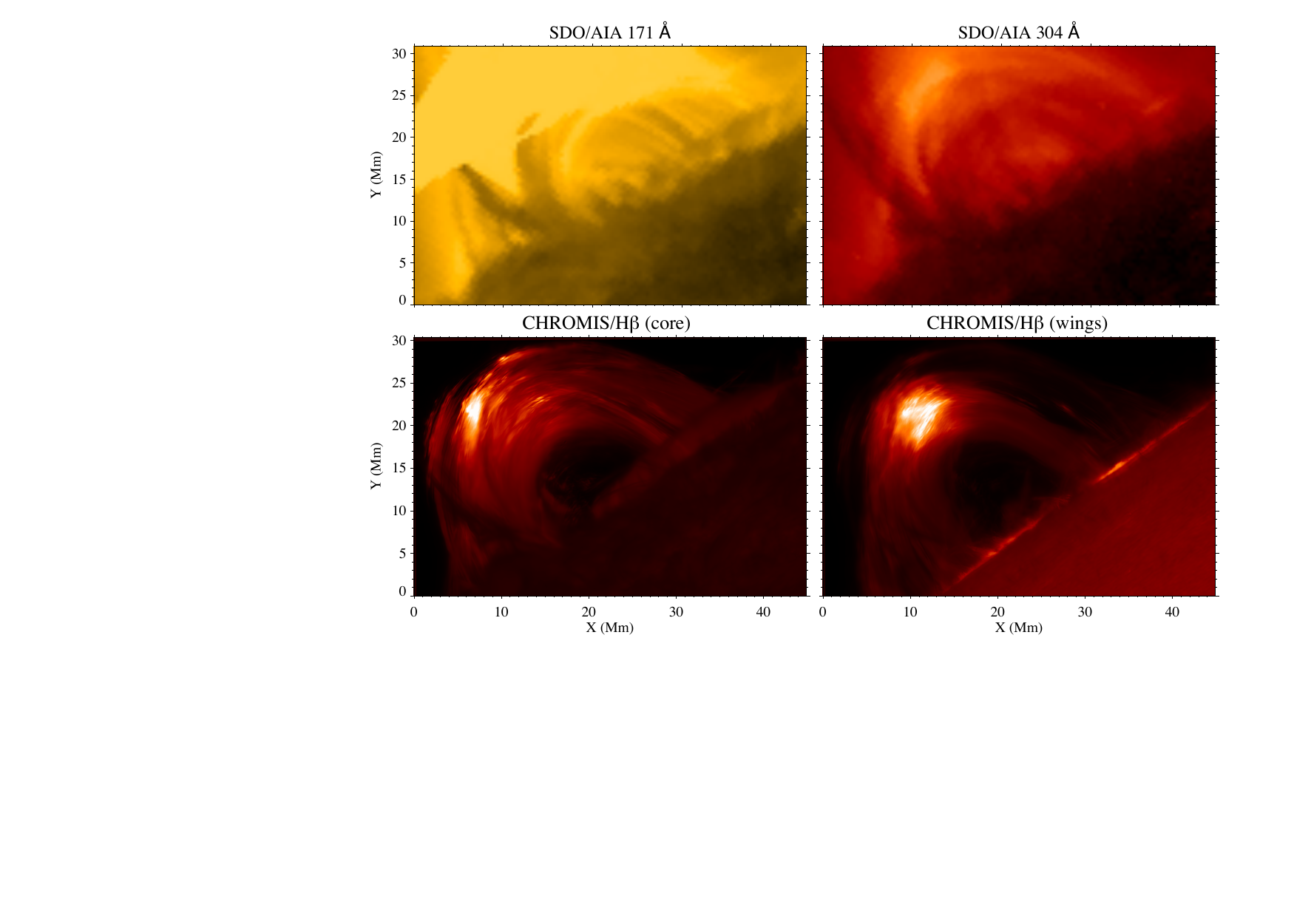}
\end{center}
\caption{SDO/AIA 171 and 304 {\AA} images (top) of the X8.2 class solar flare loops on 10 September 2017, 16:29 UT co-aligned with SST/CHROMIS H$\beta$ line core (bottom left) and the composite of H$\beta\pm0.735$ {\AA} (bottom right) images.} 
\label{fig1}
\end{figure*}

The target of our observations was NOAA active region AR12673 that produced a series of very energetic flares between 1-10 September 2017. The active region (AR) was observed between 16:07:21 and 17:58:37 on 2017 September 10 when it was close to the west limb, with heliocentric coordinates of the center of the field-of-view (FOV) at the beginning of the observations $[947", - 138"]$. 
Our observations commenced 1 minutes after the X8.2 flare peak ($\sim$16:06~UT).
The observations were carried out with the CRisp Imaging SpectroPolarimeter \citep[CRISP;][]{2006A&A...447.1111S,2008ApJ...689L..69S}  
and the CHROMospheric Imaging Spectrometer (CHROMIS) instruments, both based on dual Fabry-P\'erot interferometers (FPI) 
mounted on the Swedish 1-m Solar Telescope  \citep[SST;][]{2003SPIE.4853..341S} on La Palma. 
The imaging setup of the SST includes a dichroic beamsplitter (splits at 500 nm) and  
CRISP is mounted in the resulting red beam and CHROMIS in the blue beam \citep{2018arXiv180403030L}. 
The two instruments can collect data simultaneously but the cameras and scan sequences are not synchronized between them.
The CRISP data comprised of imaging spectropolarimetry in the Ca {\sc{ii}} 8542 {\AA} line consisted of 21 line positions with an irregular step. These positions were 
$-1.75$ {\AA} to +1.75 {\AA} ($\pm$1.75,  $\pm$0.945,   $\pm$0.735, $\pm$0.595, $\pm$0.455, $\pm$0.35, $\pm$0.28, $\pm$0.21, $\pm$0.14, $\pm$0.07,  0.0 {\AA})
from line core 
(a movie showing a 
full spectral scan is available online).
Each spectral scan of the Ca {\sc{ii}} 8542 {\AA} line had an acquisition time of 16 s but the cadence of the CRISP time series was 33 s due to inclusion of spectropolarimetric scans in the Fe {\sc{i}} 6302 {\AA}
photospheric line. However, we note that the present paper includes only the analysis of the Ca {\sc{ii}} 8542 {\AA} data, as the AR, flare loops and footpoints of these loops 
were not detected in the Fe {\sc{i}} line. The transmission FWHM for Ca 8542 {\AA} line 
is 107.3 m{\AA} with a prefilter FWHM of 9.3 \AA. Our spatial sampling was 0$''$.057 pixel$^{-1}$ over the 41$\times$41 Mm$^2$ field of view (FOV).
       
The data were reconstructed with the Multi-Object Multi-Frame Blind Deconvolution \citep[MOMFBD;][]{2002SPIE.4792..146L,2005SoPh..228..191V}. We applied the CRISP data reduction pipeline \citep{2015A&A...573A..40D} 
for standard data processing. The polarimetric calibration was performed using a linear polarizer and a quarter-wave plate at many different angles close to the primary focus on the optical table. CRISP records 4 liquid-crystal 
states per wavelength, which are linear combinations of the Stokes parameters. These states are demodulated to obtain images of the full Stokes vector $I,~Q,~U~\&~V$ components using the calibration. 
For calibrating the CRISP Ca\,{\sc II}\,8542\,\AA~data we followed the procedure described in 
\cite{2013A&A...556A.115D} and used a 
synthetic Ca\,{\sc II}\,8542\,\AA~profile computed by a 3D NLTE
radiative transfer code \citep{2011A&A...528A.113D} and a FTS atlas profile \citep{1999SoPh..184..421N} convolved with the CRISP instrumental profile. 
In order to normalize the data to the continuum intensity we used a spatially averaged Ca {\sc{ii}} 8542 {\AA} profile over the 160$\times$150 pix$^2$ 
rectangle centered at a quiet-Sun area of {\textbf(X, Y) = (34, 4)}\,Mm at $\mu \approx 0.15$ (Figure~2).


Figure 2 presents the Ca {\sc{ii}} 8542 {\AA} Stokes $I$ 
at $\mathrm{\Delta\lambda = \pm0.945~{\AA}}$
and $Q,~U,~\&~V$ monochromatic images at $\mathrm{\Delta\lambda = \pm0.455~{\AA}}$ at 16:27 UT of the flare loops.  Images show that all three polarized Stokes components are detected above the level of background emission. However, the Stokes $Q~\&~U$ profiles are very noisy and we have not used them for the measurement of the perpendicular component (with respect to the line of sight) of the magnetic field. We note that scattering polarization in Ca {\sc{ii}} 8542 {\AA} line is expecting to be very low which explains why linear polarizations detected through CRISP FPI system are weak and noisy in our data \citep{2016ApJ...826L..10S}.

We find that most of the Stokes $V$ profiles located in the lower part of the observed loop arcade 
have a signal around 10$^{-2}$, and most of the 
pixels located in the upper parts of the observed loop arcade have a signal above $\mathrm{\sim1.5\times10^{-3}}$ 
in units of continuum intensity $I_c$. These are above the noise level ($\mathrm{\sim0.5\times10^{-3}}$ at the disk and $\mathrm{\sim0.1\times10^{-3}}$ above the limb) of these Ca {\sc{ii}} 8542 {\AA} data. 

Simultaneous observations were also taken with the CHROMIS instrument. CHROMIS is a newly installed (August 2016) Fabry-P\'erot 
interferometer that observes the blue part of the spectrum in the range $3900 - 4900~\AA$. The CHROMIS observations comprised spectral imaging in 
the H$\beta$ 4861 {\AA} and Ca {\sc{ii}} H 3968.5 \& K 3933.7 $\AA$ lines, plus one position in the continuum at 4000 {\AA}. 
The Ca {\sc{ii}} H \& K line scans consist of 19 positions ranging from $-1$ {\AA} to +1 {\AA} from line core, while the H$\beta$ scan consists of 21 line positions ranging from $-1.2$ to +1.2 {\AA}. 
A full spectral scan for all three lines plus a single continuum position had a total acquisition time of 
20 s, which is the temporal cadence of the CHROMIS data time series. The transmission FWHM for the CHROMIS spectrometer \citep{2018arXiv180403030L} 
is 130 m{\AA} with a spatial sampling of 0$''$.0375 pixel$^{-1}$ and a FOV of about $\mathrm{45\times30~Mm^2}$. The CHROMIS data were processed using the CHROMISRED data reduction pipeline \citep{2018arXiv180403030L}, which includes Multi-Object Multi-Frame Blind Deconvolution image restoration. The pipeline also performs the 
calibration of the observed intensity in absolute SI intensity units by scaling the spatially averaged spectrum to an atlas profile.
                
The event was also observed with NASA's Solar Dynamics Observatory (SDO) in the Atmospheric Imaging Assembly \citep[AIA;][]{2012SoPh..275...17L} 1600/1700 {\AA} channels (Figure 3) (dominated by continuum and cool lines at $\sim$10 000 K) and several Extreme-UV channels. 
Figure 4 shows images from the Atmospheric Imaging Assembly (AIA) at 16:29 UT in the He {\sc{ii}} 304 and Fe {\sc{ix}} 171 {\AA} 
channels with emission from the transition region ($\sim$100 000 K) and corona ($\sim$1 000 000 K). 
The AIA data was recorded with a cadence of 12 seconds and spatial sampling of 0$''$.6 pixel$^{-1}$. The images are co-aligned with SST/CHROMIS data.

\section{Analysis and Results}

A well-developed set of flaring coronal loops was observed 10 minutes after the beginning of the impulsive phase of the flare (Figures 3 and 4).  
The X8.2 flare led to intense heating and evaporation of chromospheric plasma into the loop system. 
The coronal loops subsequently are filled with dense and cool plasma radiating strongly in chromospheric lines including the magnetically sensitive Ca {\sc{ii}} 8542 {\AA} (Figures 2, 3, and 4).
We note that the origin of the cool plasma in flare coronal loops is still under debate in the solar community. 
However, rapid cooling of evaporated plasma through radiation losses is considered as the most plausible explanation of these phenomenon \citep{1987ApJ...320..409K,1996SoPh..165..303S}.
Unfortunately, footpoints of the observed flare loops were not detected in SST data suggesting that they are behind the limb.

The off-limb location of the flare coronal loops minimized contaminations in the chromospheric spectra caused by line-of-sight on-disk effects, providing polarisation data of unprecedented quality.

\subsection{LOS Magnetic field}

\begin{figure}[t]
\begin{center}
\includegraphics[width=8.8cm]{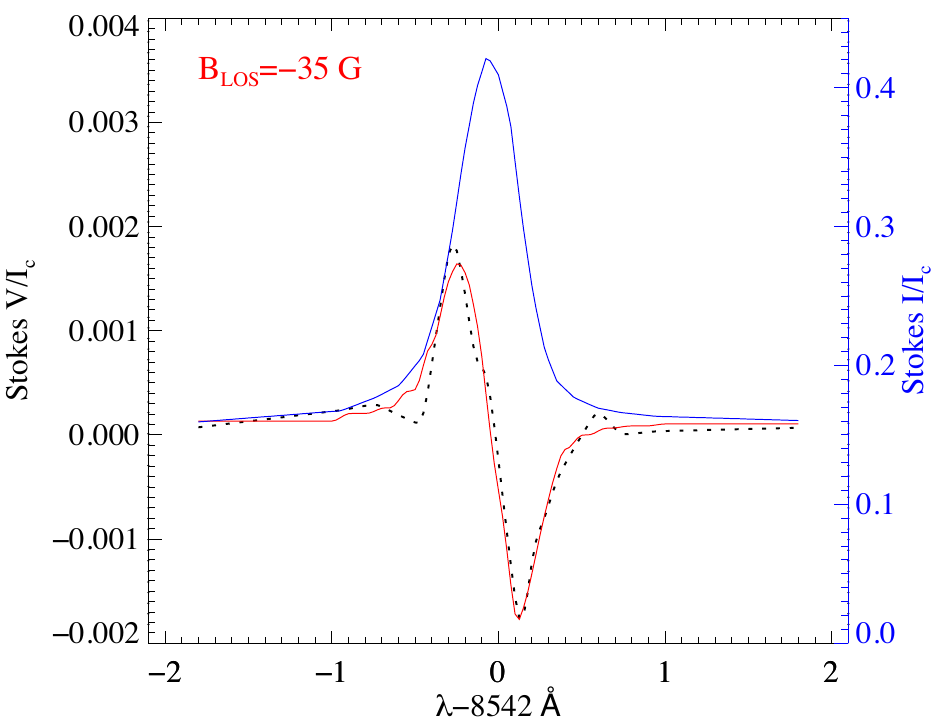}
\end{center}
\caption{Circular polarization profile (black dashed line) of the pixel with the weakest LOS magnetic field (X=15 Mm, Y=30 Mm in Fig. 2) at 16:28 UT. The blue line shows the intensity (Stokes $I$), and the solid red line shows the WFA fit obtained from the derivative of Stokes $I$.} 
\label{fig1}
\end{figure}

\subsubsection{Weak-field approximation}

Circular polarisation images (Stokes $V$) of the observed flaring loops acquired in Ca {\sc{ii}} 8542 {\AA} reveal strong polarization signals along the loops (Figure 2). 
We use the weak field approximation (WFA) technique to produce maps of the line-of-sight (LOS) magnetic field component ($B_{LOS}$) for the flare coronal loops. The WFA is the simplest approach to calculate the magnetic flux density from the observed line intensity and polarization \citep{2004ASSL..307.....L,1993SoPh..146..207C,2017ApJ...834...26K,2013A&A...556A.115D,2011ApJ...731...27A,2018ApJ...866...89C,2018ApJ...860...10K}. More complex methods such as inversions of polarimetric data are not applicable in this case as they are not optimized for off-limb chromospheric/coronal spectropolarimetric observations. The WFA can provide an accurate estimate of the chromospheric and coronal magnetic field. The main limitation of the WFA is that it can be applied only if the Zeeman splitting ($\Delta\lambda_H$) 
is much smaller than the Doppler width ($\Delta\lambda_D)$ of the observed spectral line 
(Ca {\sc{ii}} 8542 {\AA} line in our case)  \citep{2013A&A...556A.115D},
\begin{equation}
\Delta\lambda_H\ll\Delta\lambda_D.
\end{equation}
Furthermore, the magnetic field and LOS velocity have to be close to a constant as a function of distance along the LOS in the atmosphere. In this regime the Stokes profiles can be expressed as \citep{1977A&A....56..111L},   
\begin{equation}
V(\lambda)= - 4.67\times10^{-13}g_{eff}\lambda_0^2B_{LOS}\frac{\partial I(\lambda)}{\partial \lambda},
\end{equation}
\begin{equation}
\left[Q^2+U^2\right]^{1/2}= \left |- 5.45\times10^{-26}\bar{G}\lambda_0^4B^2_{\bot}\frac{\partial^2 I(\lambda)}{\partial \lambda^2}\right |,
\end{equation}
where $g_{eff}$ is the first order effective Land\'e factor,  
and $\lambda_0$ the central wavelength of the spectral line.
$\bar{G}$ is a second order effective Land\'e factor related to $g_{eff}$ with
\begin{equation}
\bar{G}=g_{eff}^2-(g_1-g_2)^2(16s-7d^2-4)/80,
\end{equation}
where
$$
s=J_1(J_1+1)+J_2(J_2+1),~~d=J_1(J_1+1)-J_2(J_2+1)
$$
for the angular momentum $J_1$ and $J_2$ of the involved energy levels with Land\'e factors  $g_1$ and $g_2$ \citep{1977A&A....56..111L}.
For Ca {\sc{ii}} 8542 {\AA} line $g_{eff}$=1.1 and $\bar{G}$ =1.21. We note that  the units for the
wavelength and magnetic field are {\AA} and G, respectively.   

\begin{figure}[t]
\begin{center}
\includegraphics[width=8.65cm]{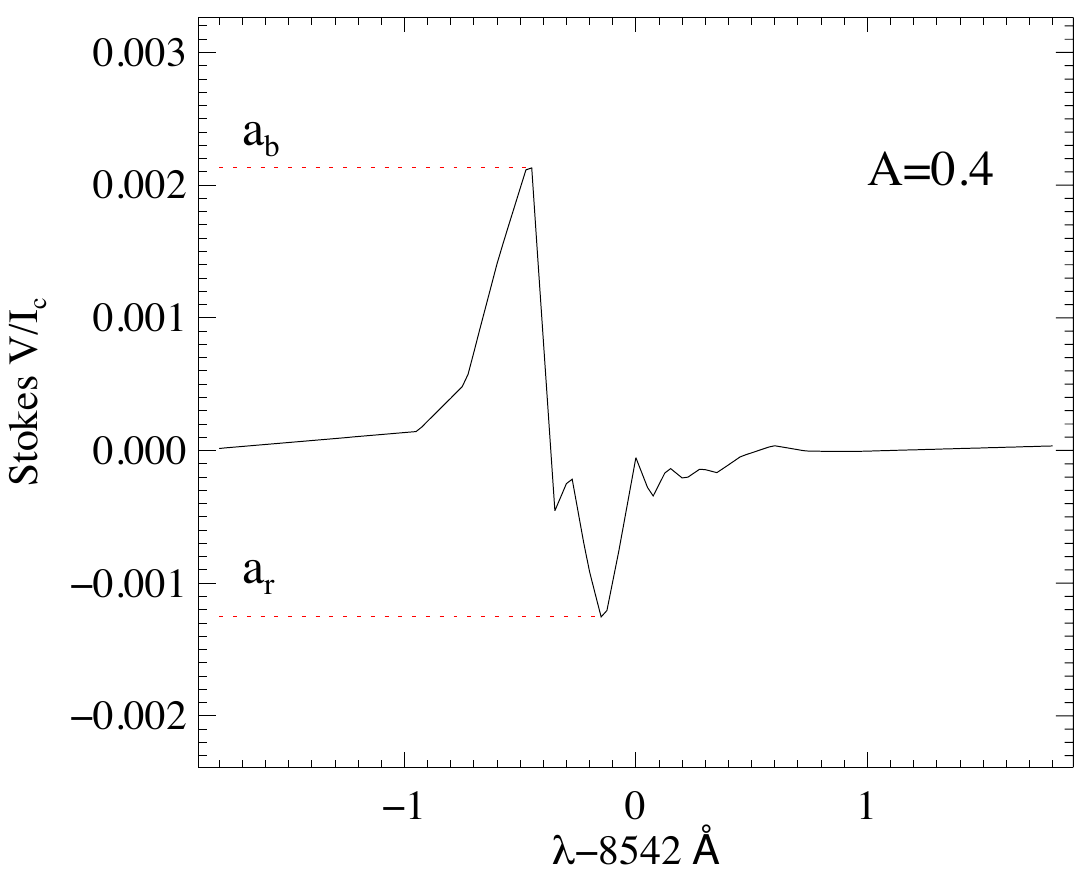}
\end{center}
\caption{Asymmetric Stokes $V$ profile observed for a pixel located at around X=16 Mm, Y=32 Mm (Fig. 2). $a_b$ and $a_r$ denote the red and blue wing amplitudes which define the amplitude asymmetry, $A$.} 
\label{fig1}
\end{figure}

\begin{figure*}[t]
\begin{center}
\includegraphics[width=12.9cm]{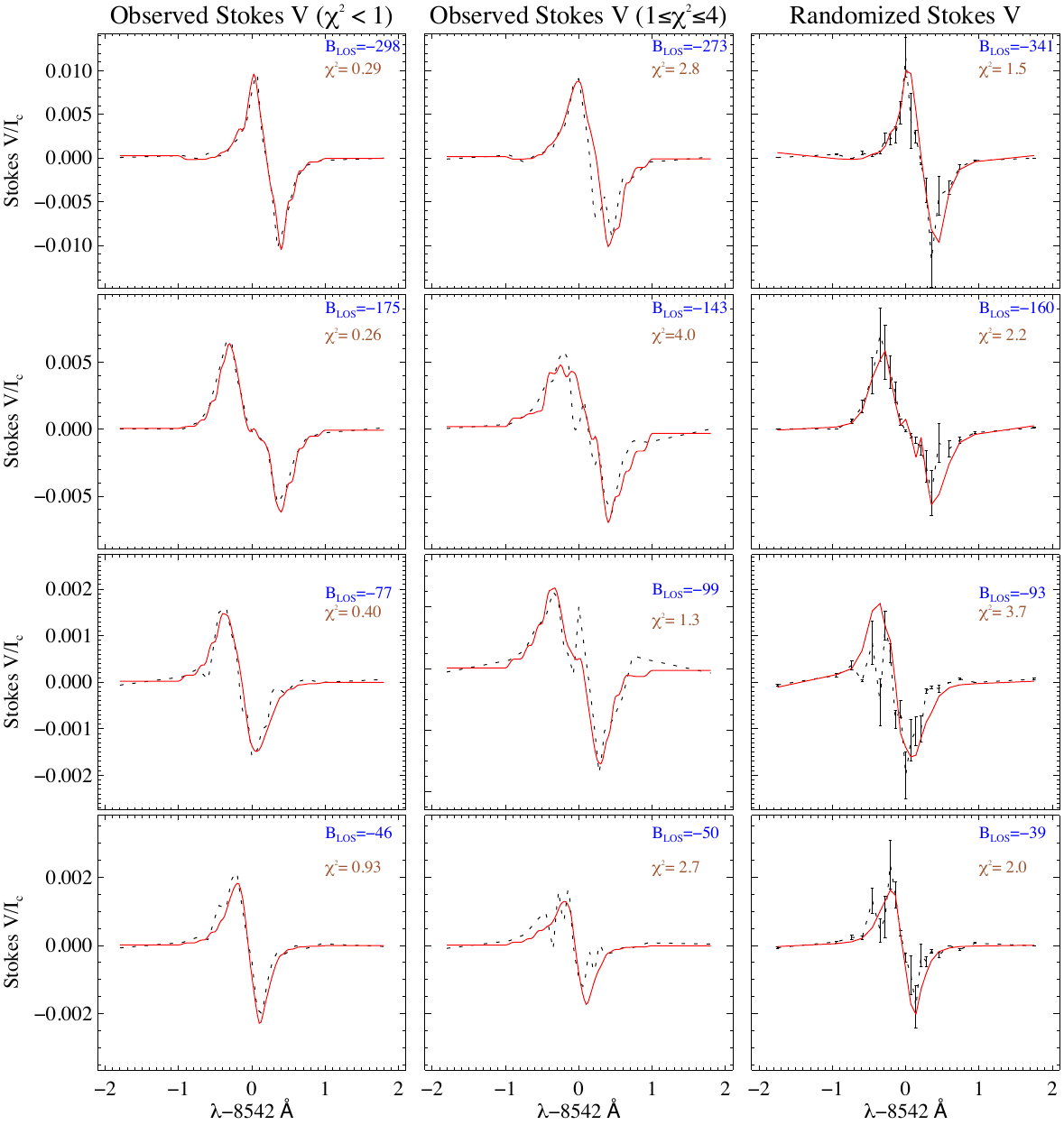}
\includegraphics[width=5.03cm]{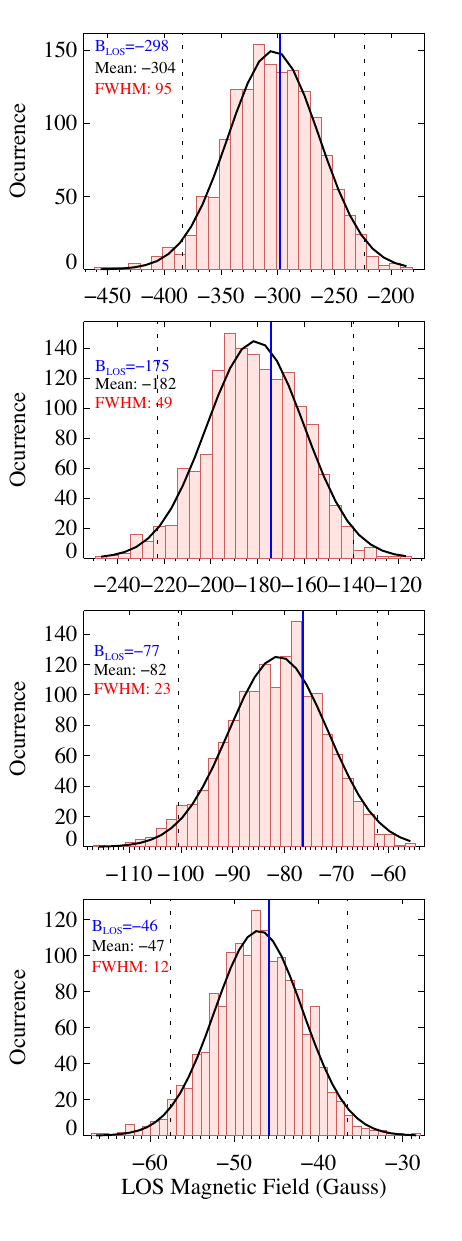}
\end{center}
\caption{Column 1: Observed Stokes {\it V} profiles (black dashed lines) in the pixels with the lowest noise level and goodness-of-fit  $\chi^2<1$. The WFA fits obtained from the derivative of Stokes I are depicted as solid red lines. Column 2: Typical Stokes profiles with higher noise level and goodness-of-fit with $1\leqslant\chi^2\leqslant4$. Column 3: Randomized Stokes {\it V} profiles presented in the column 1. 
The randomization degrades their WFA goodness of fit to $1\leqslant\chi^2\leqslant4$. Column 4: Histograms of the magnetic field strength values derived with the WFA from the derivative of the randomized Stokes {\it I} and Stokes {\it V} presented in the third column. The vertical blue solid line on each histogram indicates the values of $B_{LOS}$ computed with the WFA before randomization of the Stokes  {\it V~\&~I}. Error bars represent standard deviation values from the 1500 randomisation. Gaussian fits of the histograms are presented as the full black lines with the 95$\%$ confidence intervals (the vertical dashed lines).} 
\label{fig1}
\end{figure*}

The low Land\'e factor and relatively broader line width due to the increased temperature (compared to photospheric temperature) of Ca {\sc{ii}} 8542 {\AA} 
suggests that the WFA is an appropriate  method for this line. 
Under chromospheric conditions the upper limit of the WFA for the Ca {\sc{ii}} 8542 {\AA} is estimated approximately $B\leqslant2500$ G \citep{2013A&A...556A.115D,2011ApJ...731...27A}.
The average field strength in the chromosphere and corona 
are well below this value so the WFA  should be valid method 
for inferring the magnetic field using the Ca {\sc{ii}} 8542 {\AA} line.
For most of the pixels in the observed coronal loops Equation (1) is fulfilled. Line profiles, which have not met the criterion presented in Equation (1) are excluded from the analysis. 
Figure 5 shows an example of the Stokes $I \& V$ profiles, and the WFA fit obtained from the derivative of Stokes $I$, for the pixel with the weakest LOS magnetic field.

As it was mentioned above the WFA is applicable when the magnetic field and velocities are close to a constant along the LOS. The 
Ca {\sc{ii}} 8542 
line emission is formed over a wide layer of the solar atmosphere suggesting that due to the vertical stratifications the magnetic field can have a strong gradient as a function of height in the atmosphere. However, off-limb coronal loop observations provide less inhomogeneity along the LOS.  Magnetic field/velocity gradients and/or discontinuities can produce asymmetric Stokes $V$ profiles \citep{2003A&A...408.1115K}.
Therefore, asymmetries can be used as a good indicator of magnetic field/velocity gradients.

To quantify the asymmetry of the Ca {\sc{ii}} 8542 line Stokes $V$ profiles
we use a technique similar to that described in \cite{2001ApJ...563.1031S}.
We measured the absolute values of amplitudes of the red and blue lobes, $a_b$ and $a_r$, respectively for every pixel in the observed coronal loops 
and compute amplitude asymmetry with  
\begin{equation} 
A=\frac{\left| a_b-a_r \right|}{max(a_r,a_b)}.
\end{equation}
$A=0$ corresponds to no amplitude asymmetries in the Stokes {\it V} line profiles and $A=1$ corresponds to maximum asymmetry which is the case when Stokes {\it V} has a single lobe profile.  Figure 6 shows an asymmetric Stokes $V$ profile with $A=0.4$. Profiles with asymmetry higher than 0.4 were excluded from the analysis. This value was chosen based on a statistical analysis of randomized profiles (see section \ref{unc}) indicating that the asymmetries below this value are very likely introduced by the noise rather than by the gradients in magnetic field and velocity (Figure 7).

\begin{figure*}[t]
\begin{center}
\includegraphics[width=18.12cm]{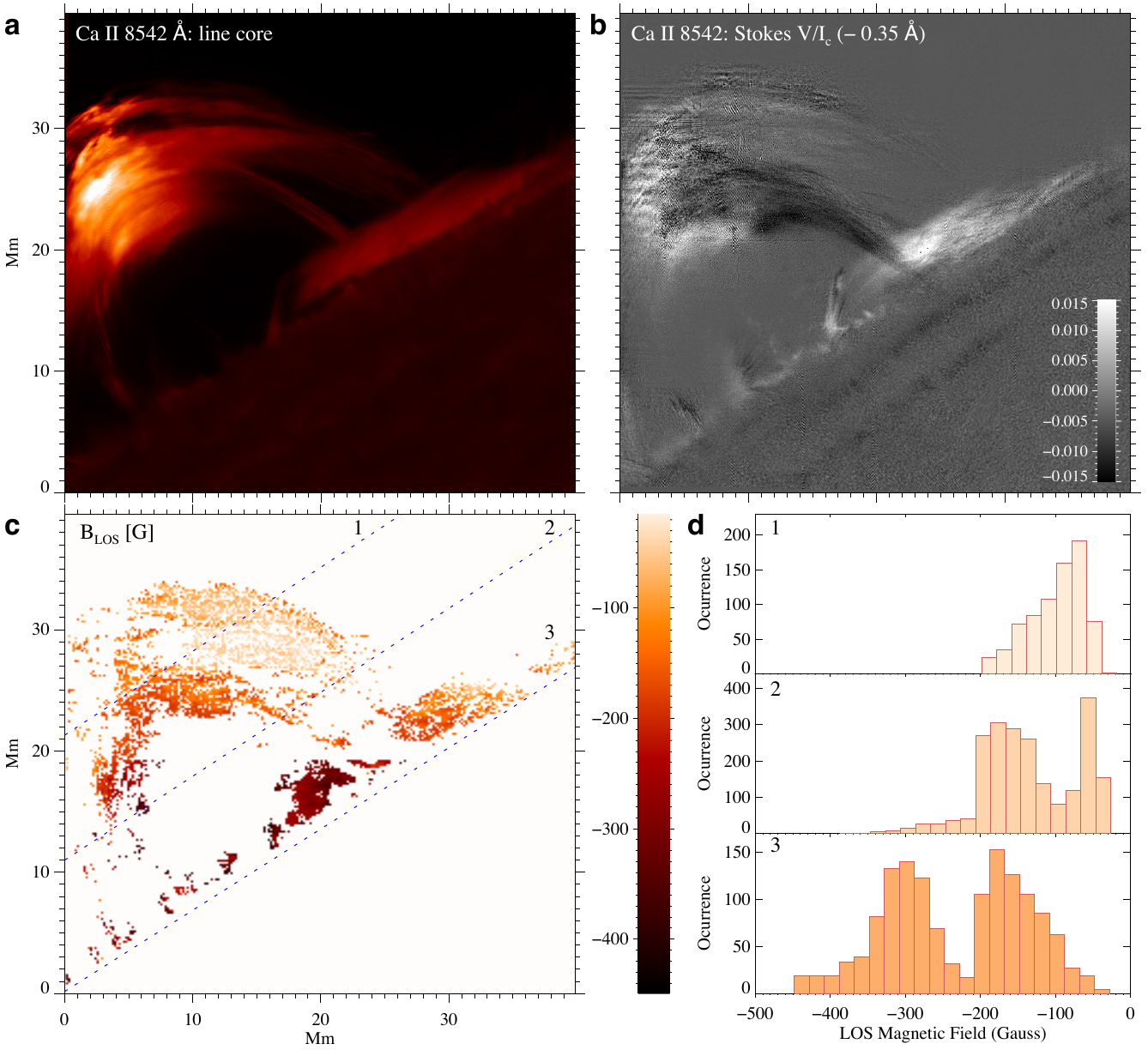}
\end{center}
\caption{Panel $a$ and $b$ show the SST images in the Ca {\sc{ii}} 8542 {\AA} intensity (Stokes {\it I}) at line core and circular polarization (Stokes {\it V}) 
at line wing for the flare coronal loops at 16:28 UT, 22 minutes after the flare peak. Panel $c$ and $d$ show a map of the LOS magnetic 
field together with histogram showing the distribution of the LOS magnetic fields for 3 different regions $(1, 2, 3)$.} 
\label{fig1}
\end{figure*}

Equation 2 shows that $\partial I(\lambda)/\partial \lambda$ can be used as a calibration constant for LOS magnetic field component.
To extend the wings of the observed Stokes profiles we added two more wavelength points at $\pm$1.8 {\AA} 
and extrapolate the profiles at these points.
Stokes profiles were also averaged 
over $0.23''\times0.23''$ ($\sim$170~km) area in order to improve S/N ratio. 
The derivatives of Stokes $I$ were then calculated with IDL's built-in deriv() function for every pixel of the observed coronal loops.
Then the best fit between the $V(\lambda)$ and $\partial I/\partial \lambda$ is computed with the least square minimisation technique.
The fitting coefficient directly gives the values for the magnetic field.
The parameter $\chi^2$ that characterizes the discrepancies (goodness-of-fit) between Stokes $V$ and the derivatives was also computed for every pixel. Figure 7 (first and second columns) shows  Stokes $V$ profiles from different parts of the observed coronal loops, together with the WFA fits obtained from the derivative of Stokes $I$ (overplotted as solid red lines). The first column of Figure 7 shows the results for lower $\chi^2$ ($<$1) and the second column shows the results with higher $\chi^2$ ($\sim$1-4).

Figure 8c shows the resulting map of the LOS magnetic field, across large regions of the loop system. 
The histograms in Figure 8d are the distributions of the LOS magnetic field for 3 different height ranges. $B_{LOS}$ of the loop apex region 
(layer 1, between 18 and 26 Mm above the solar surface) ranges from 50 to 180 G with median 90 G. 
The corresponding values at mid-heights (layer 2, 9 to 18 Mm) are as high as 300 G with a median of 140 G. 
The polarization signal of the loop legs in the lower part of the FOV (layer 3, up to 9 Mm) is very weak due to the increased 
perpendicular orientation of the loop legs. Layer 3 is also contaminated by a solar prominence associated with the flare, 
lying along the LOS. The distribution of LOS magnetic field here shows two distinct peaks at 150 and 300 G.

Compared to the LOS component, it is more challenging to compute the perpendicular component of the magnetic field with the WFA fitting as it is related to the total linear polarization, which depends on the second derivative of Stokes $I$ with respect to wavelength (Equation 3). The measured linear polarization signals (Stokes $Q~\&~U$) are noisy and could not be used to estimate the component of the magnetic field that is perpendicular to our LOS under the WFA limit (Equation 3).

\subsubsection{Influence of spectral sampling}
\label{inf_samp}

\begin{figure*}[t]
\includegraphics[width=6.cm]{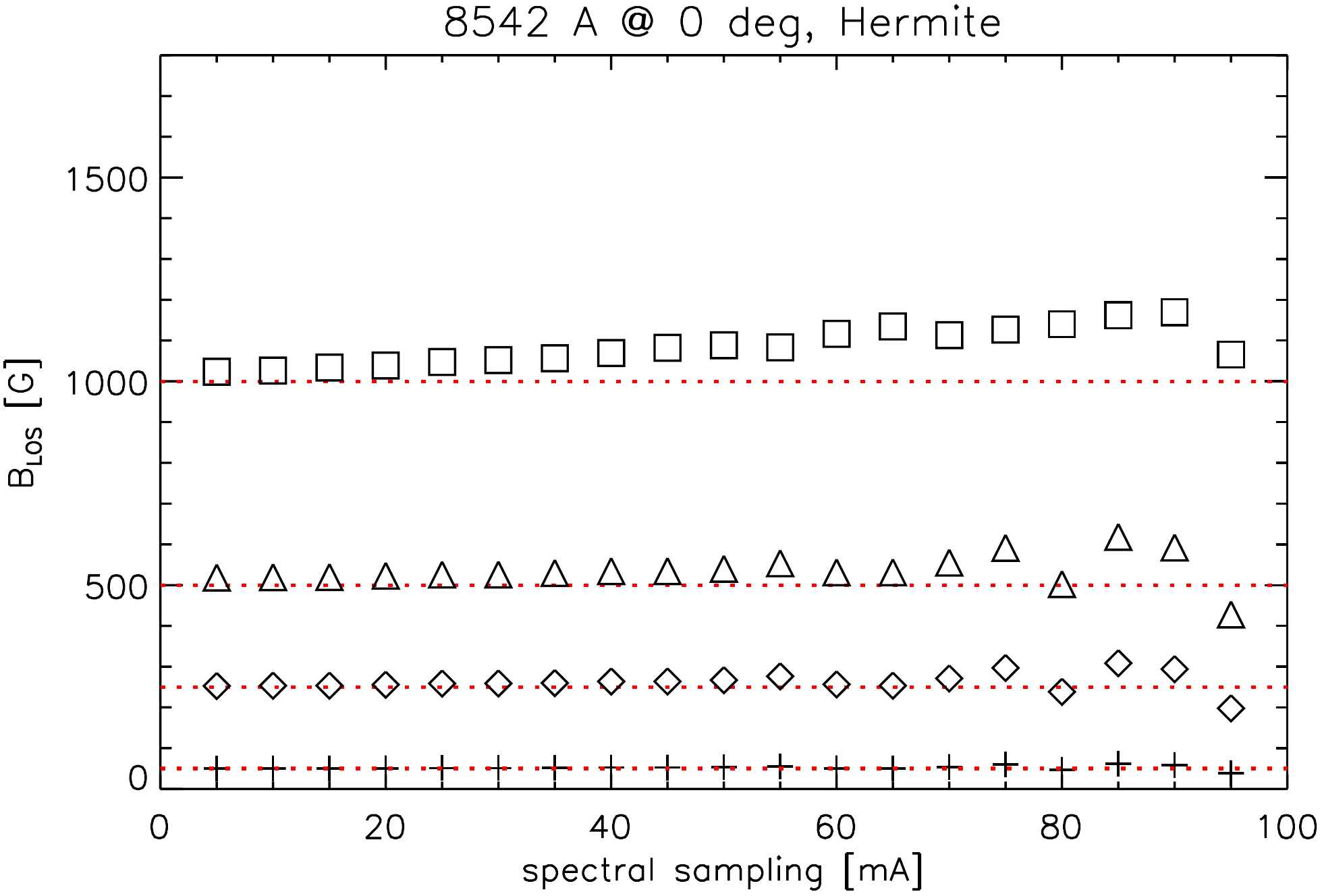}
\includegraphics[width=6.cm]{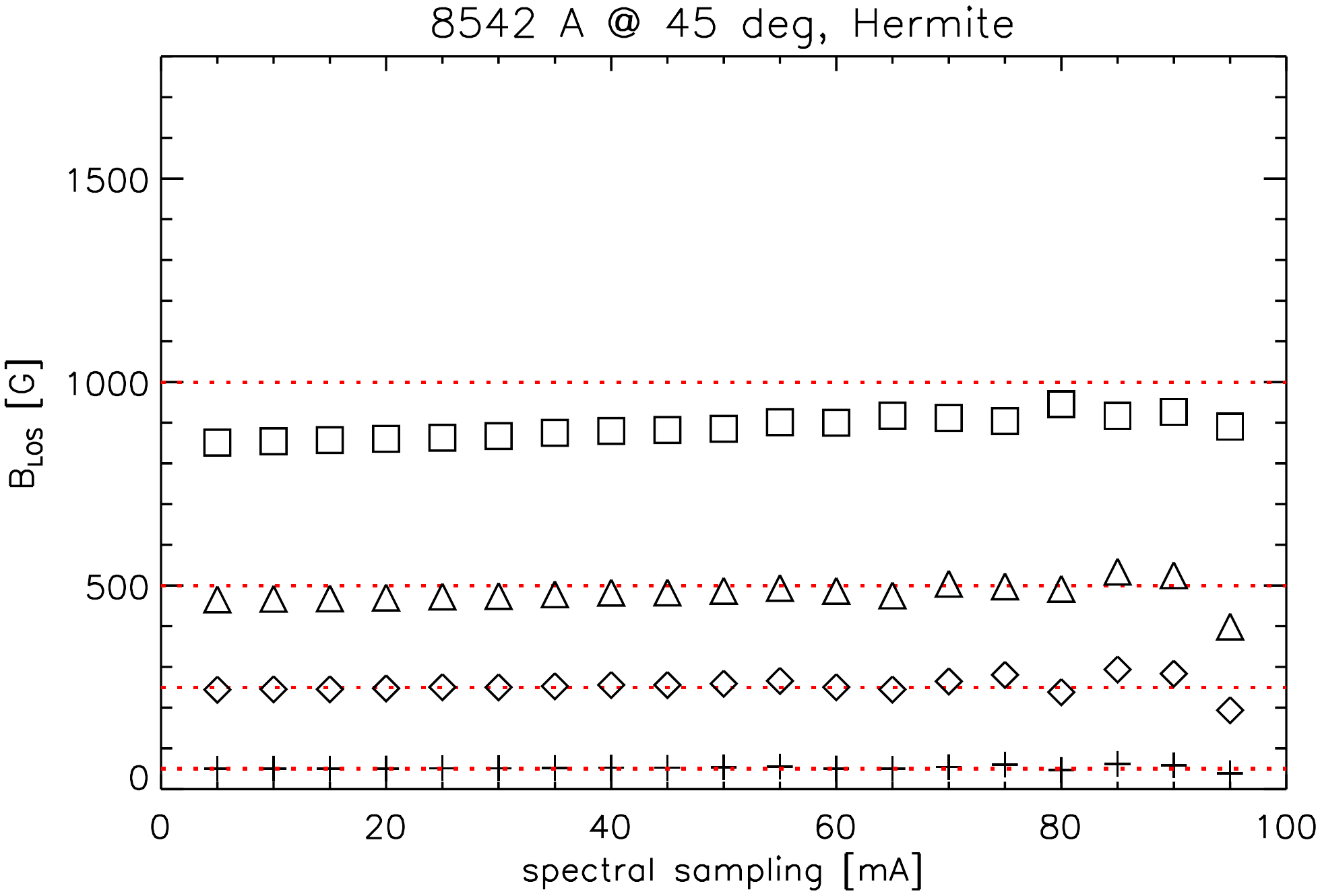}
\includegraphics[width=6.cm]{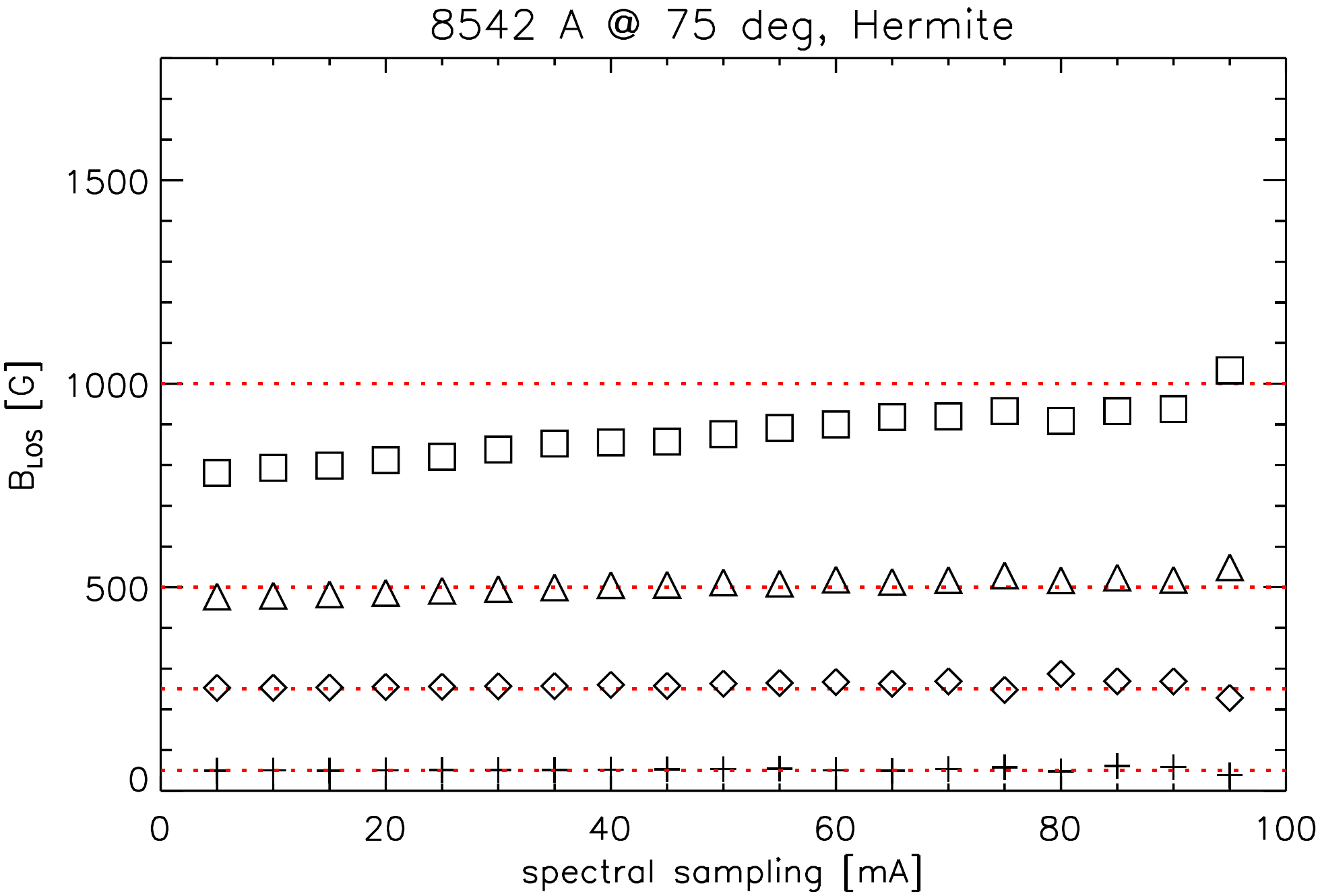}
\caption{Magnetic field values derived from the WFA for different spectral samplings for the synthetic Ca {\sc{ii}} 8542 {\AA} data. 
The input fields are denoted by horizontal dotted lines and the calculated values with the WFA are given by 
the various symbols (+: 50 G, diamond: 250 G, triangle: 500 G, box: 1000 G). The left panel shows a purely 
vertical magnetic field, the middle panel a field with inclination 45$^\circ$ and the right 
panel a field with inclination 75$^\circ$ with respect to the solar surface, which influences the errors of the derived values.} 
\label{fig1}
\end{figure*}

\begin{figure*}[t]
\begin{center}
\includegraphics[width=14.5cm]{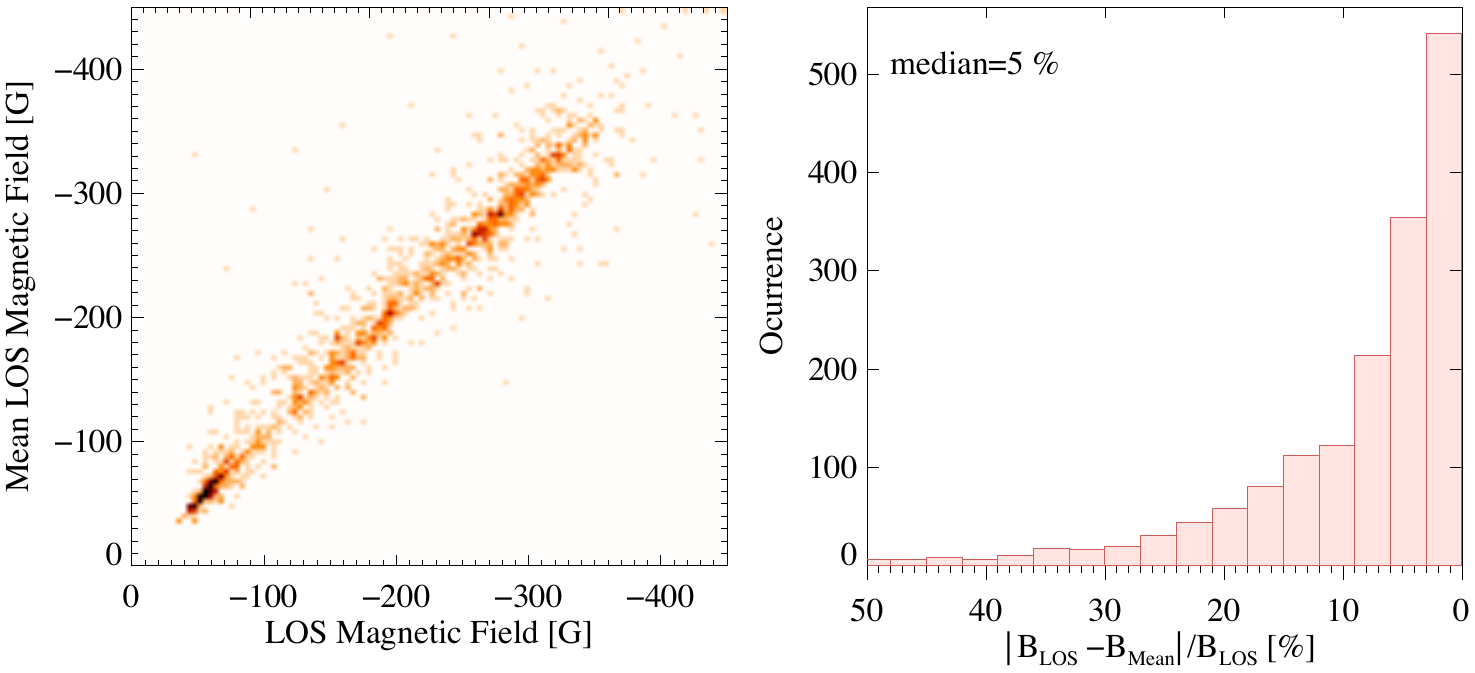}
\end{center}
\caption{Scatter plot represented as a two-dimensional density map
of the LOS magnetic field measured from the WFA of the initial, high quality Stokes profiles versus the mean $B_{LOS}$ retrieved after the randomization test.}
\label{fig1}
\end{figure*}

Kleint \& Sainz Dalda 2015 (private communication) have investigated the influence of the spectral sampling on the 
WFA for the Ca {\sc{ii}} 8542 {\AA} line by synthesizing the line in Non-Local Thermodynamical Equilibrium (NLTE), with the NICOLE inversion code \citep{2015A&A...577A...7S}. 
They used different input values (between 50 - 1000 G) for the LOS magnetic field and computed synthetic profiles for different spectral sampling (from 5 to 100 m{\AA}), model atmospheres (VAL-C and HSRA), and predefined inclinations. The $B_{LOS}$ was then computed with the WFA and the results were compared to the original input magnetic field values to estimate the influence of spectral sampling on the WFA. Figure 9 shows that the difference between input values and derived magnetic fields generally increases when the spectral sampling increases for a kG field. 
However, for fields $\leqslant$500 G there is a very good match between them even for large spectral sampling. 
The $B_{LOS}$ derived in the present work with the WFA are below 400 G and the spectral sampling of our data near the line core is 70 m$\AA$. This suggests that in locations with well-defined Stokes profiles with a good fit between Stokes {\it V} and the derivative of Stokes {\it I}, the WFA provides a very accurate measure of the magnetic field density, with error smaller than 10$\%$.

\subsubsection{Uncertainty of the measurement}
\label{unc}

\begin{figure*}[t]
\begin{center}
\includegraphics[width=18.1cm]{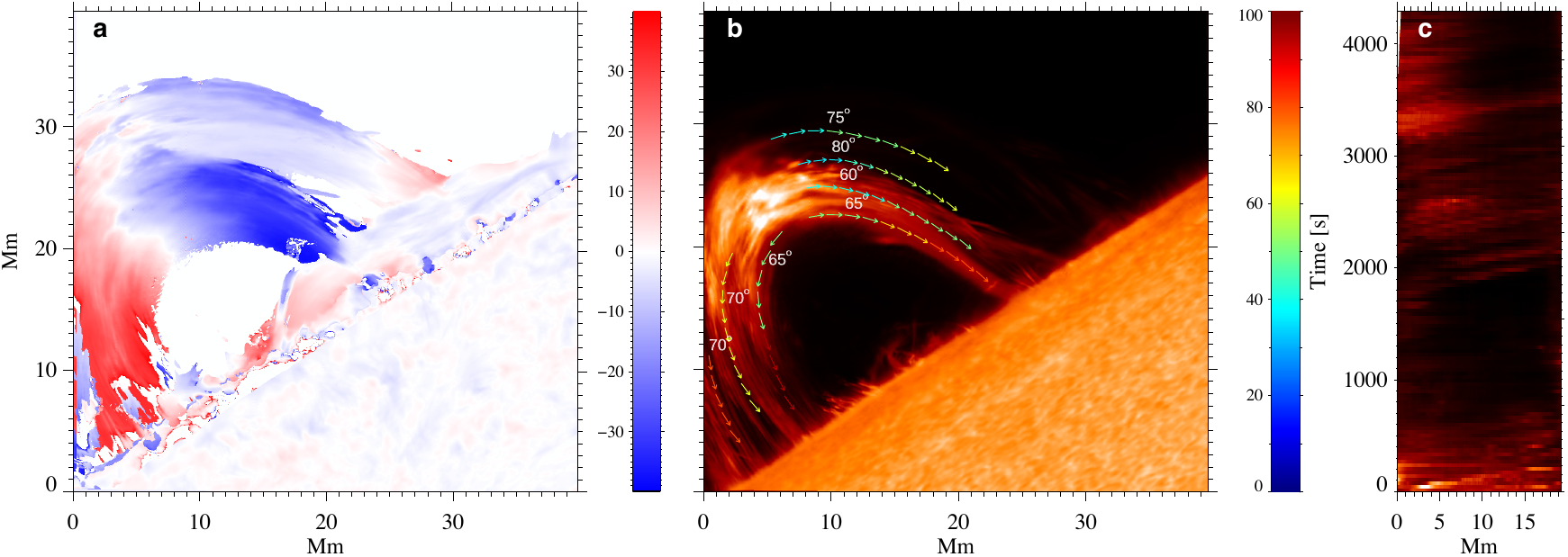}
\end{center}
\caption{(a) Doppler map of the flaring coronal loops. (b) Composite of CRISP Ca {\sc{ii}} 8542 $\pm$ 0.945 {\AA} images. 
The color-coded arrows indicate direction and speed of plasma flows. The numbers above the arrows indicate average viewing 
angles of the flow/magnetic loops with respect to the LOS direction. The colour bars give velocity in $\mathrm{km~s^{-1}}$. 
(c) Time-distance diagram of Ca 8542 $\pm$ 0.945 {\AA} intensity along the outermost arrows in panel $b$.} 
\label{fig1}
\end{figure*}

\begin{figure}[t]
\begin{center}
\includegraphics[width=9.1cm]{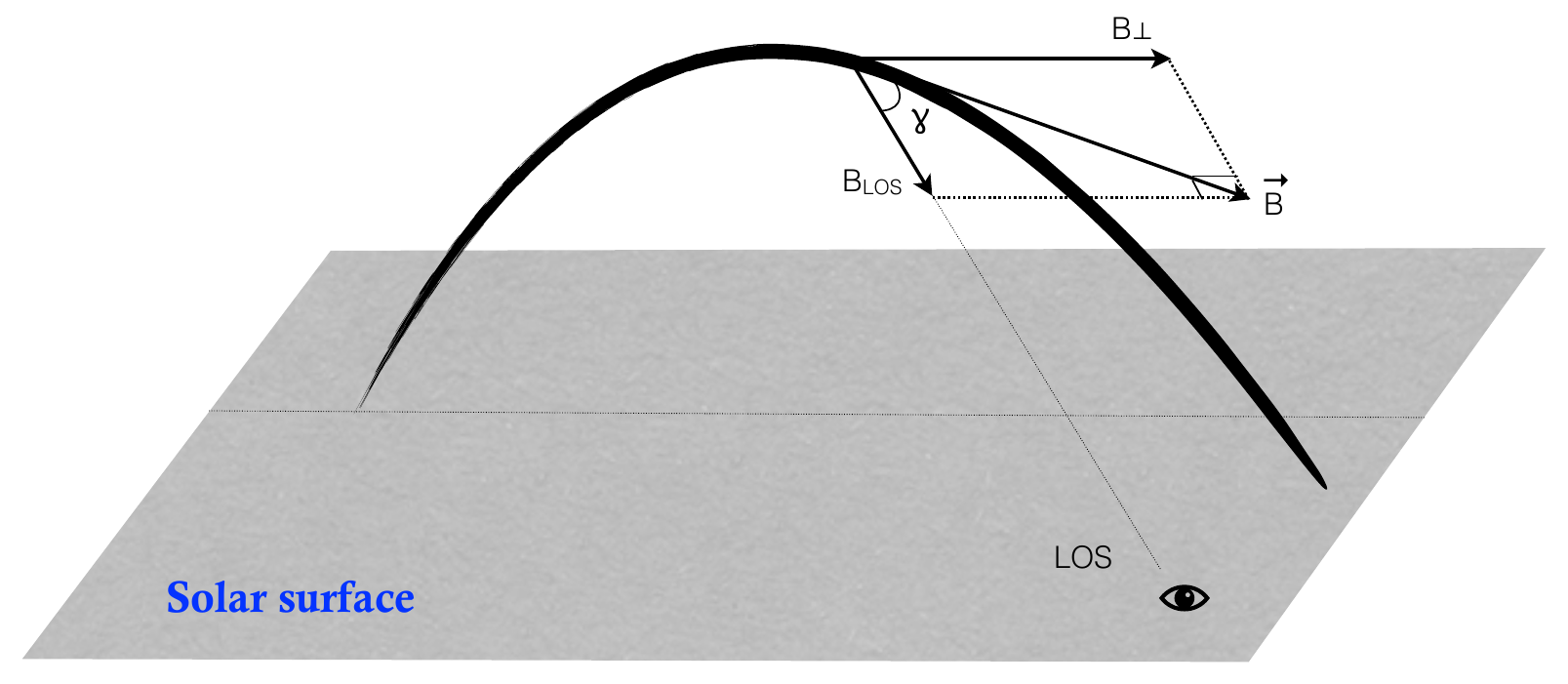}
\end{center}
\caption{Schematic representation of the magnetic loop showing the basic geometry, magnetic field vector and its LOS and perpendicular components with respect to the observer.} 
\label{fig1}
\end{figure}

A large fraction (20\%) of the pixels selected for this analysis and presented in the LOS magnetic map (Figure 8c) have well-defined, regular, symmetric Stokes $V$ profiles with 
$\chi^2\leqslant1$. Examples of such profiles are presented in Figure 5 and the first column of Figure 7. As it was discussed in section~\ref{inf_samp}, for such profiles the estimated error of the measured magnetic field strength is less than 10\%. Typical profiles with higher noise level and hence higher goodness of fit ($1\leqslant\chi^2\leqslant4$) for the WFA are shown in the second column of Figure 7. Profiles with $\chi^2 > 4$ were ignored and are not included in the magnetic field map (Figure 8). To estimate an influence of the noise on the $B_{LOS}$ values derived by the WFA we performed randomization tests. The method has been applied in the following way:  we chose the 3 best scans from our time series in terms of spatial resolution. From each selected scan we chose the pixels with the Stokes $V~\&~I$ profiles with the lowest noise level and lowest WFA goodness-of-fit ($\chi^2\leqslant1$). For these pixels we applied artificial noise by using a randomization of the Stokes $V~\&~I$ profiles 
and degraded their WFA goodness of fit to $1\leqslant\chi^2\leqslant4$. The third column of Figure 7 shows the Stokes $V$ profiles produced by such randomization from the profiles presented in the first column of Figure 7. The process is repeated 1500 times for each selected pixel, producing a distribution of values for the LOS magnetic field. The right column of Figure 7 shows the histograms of the magnetic field values derived with the WFA after randomization of Stokes $V$ profiles with the lowest noise level (presented in the left column of Figure 7). The vertical solid blue lines on each histogram indicate the values of $B_{LOS}$ computed with the WFA before randomization of the Stokes $V~\&~I$. 

The left panel of Figure 10 shows the density map of the LOS magnetic field measured from the WFA of the initial, high quality Stokes profiles versus the mean, $B_{Mean}$ retrieved after the randomization test. 
The right panel of Figure 10 shows the relative error of the $B_{Mean}$ 
compared to the original $B_{LOS}$ obtained from the WFA is around 5$\%$. 
This suggests that the upper limit of the uncertainty of the LOS magnetic field derived from the WFA for the profiles with goodness-of-fit $1\leqslant\chi^2\leqslant4$ 
should be close to the 2$\sigma$ uncertainty ranges, which is less than 30\% of $B_{Mean}$ (Figure 7).

\subsection{Velocity map}

\begin{figure*}[t]
\begin{center}
\includegraphics[width=17.3cm]{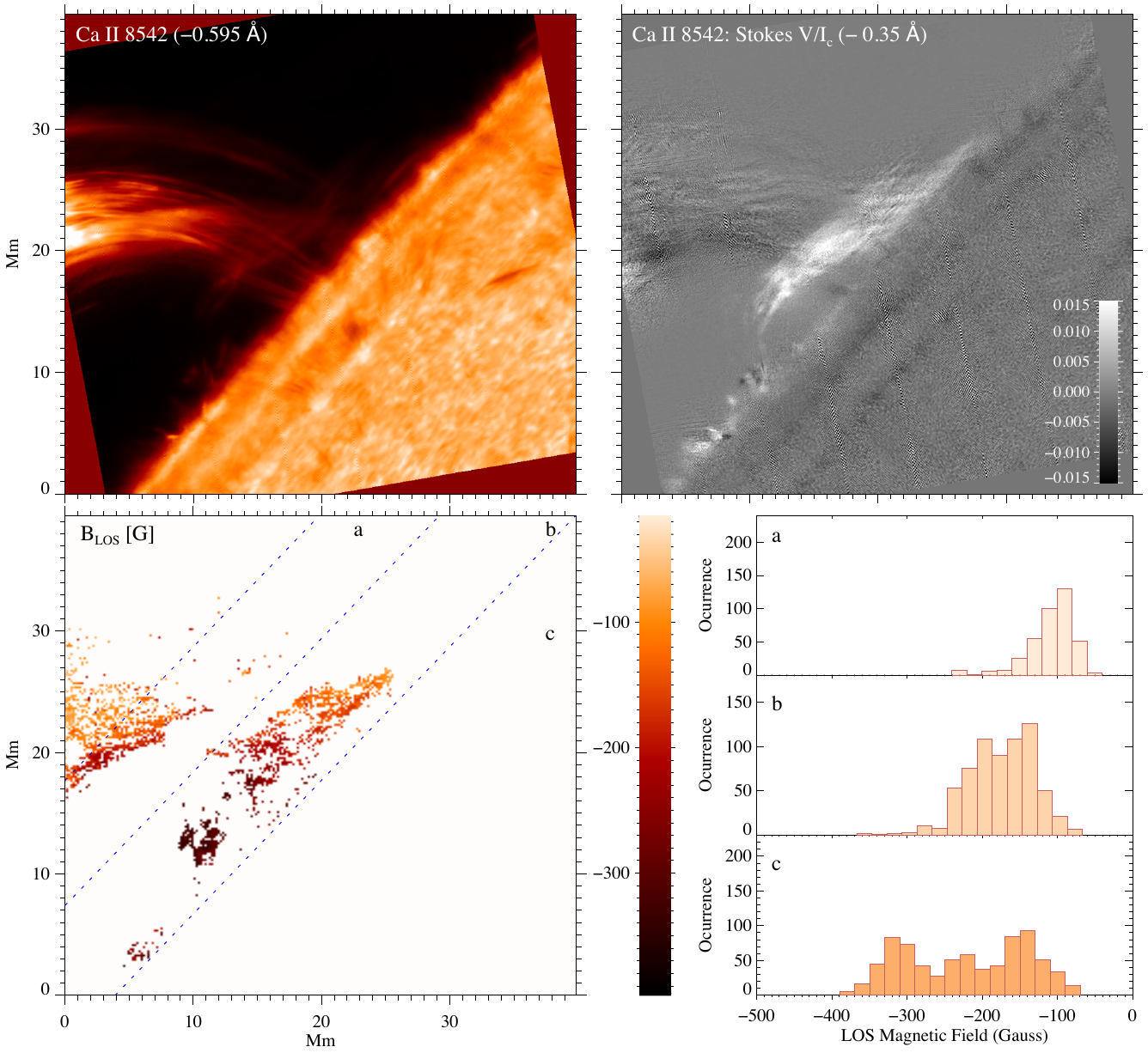}
\end{center}
\caption{Same as Figure 8 obtained 5 min later at 16:33 UT.} 
\label{fig1}
\end{figure*}

\begin{figure*}[t]
\begin{center}
\includegraphics[width=17.3cm]{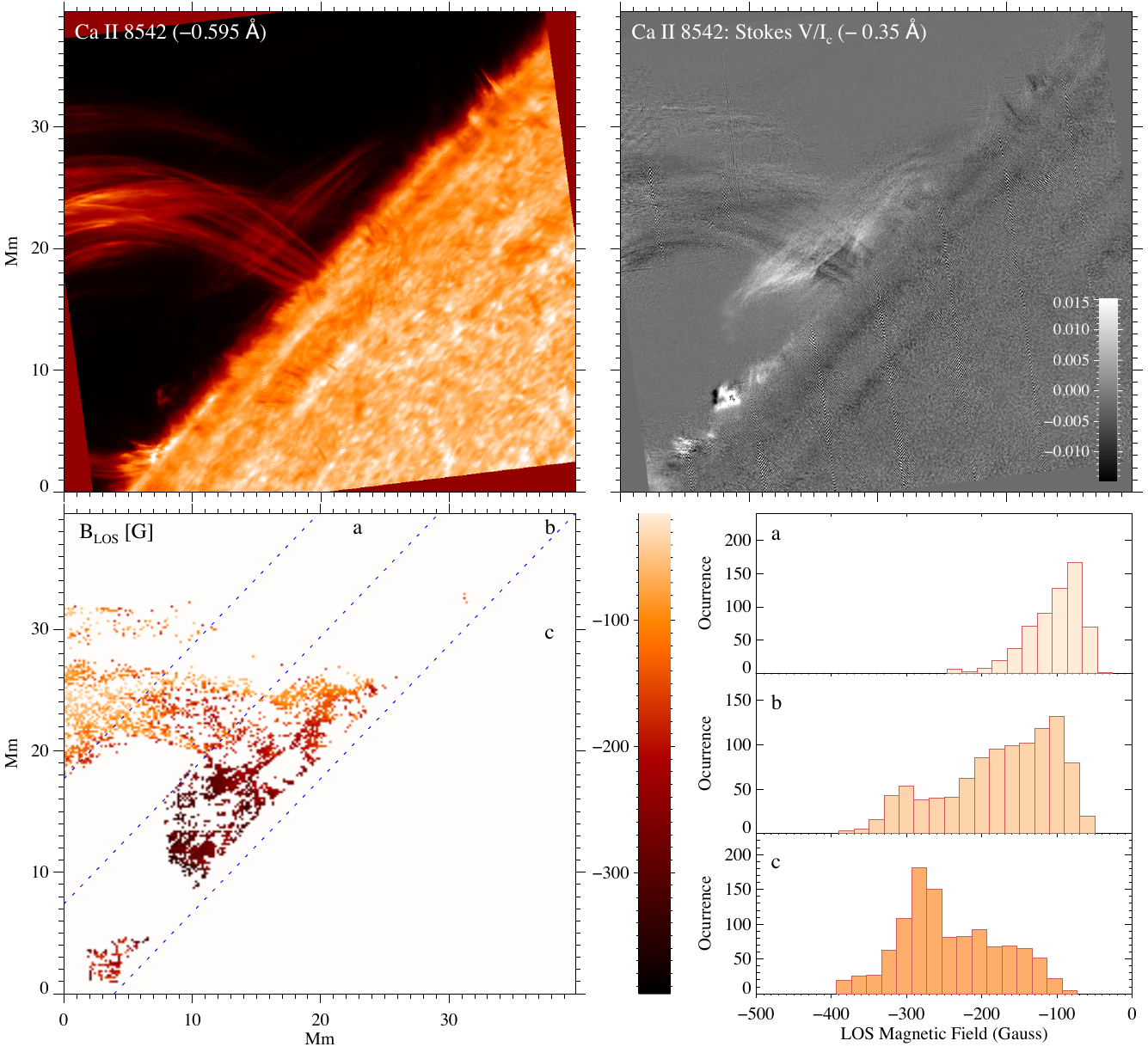}
\end{center}
\caption{Same as Figures 8 and 13 just at 16:45 UT.} 
\label{fig1}
\end{figure*}

The line profiles are fitted with Gaussian functions to determine the bulk plasma LOS velocity (Figure 11a). 
Most of the CRISP/Ca {\sc{ii}} 8542 {\AA} line profiles are well-defined, single peaked emission profiles. However, many pixels ($\sim$40\%) 
of the coronal loops in the CHROMIS/Ca K \& H and H$\beta$ lines have irregular, noisy profiles and they cannot be fitted reliably with a single Gaussian. 
As CHROMIS covers the shorter wavelengths it is more susceptible to seeing conditions compared to 
CRISP and changes in atmospheric seeing during the scan time of spectral lines has affected the 
quality of some of the CHROMIS data. The Doppler velocities measured for the pixels with the well-defined profiles in CHROMIS 
lines are consistent with the velocities obtained from the CRISP Ca {\sc{ii}} 8542 {\AA} line indicating the reliability of the measurements.
Mapping this value (Dopplergrams) reveals that the left and right part of the loop structures contain regions of red- and blue-shifted profiles, respectively, with velocities between $\mathrm{\sim10 - 35~km~s^{-1}}$ (Figure 11a). 
Time-distance diagrams show strong {gravity-driven} downflows of dense and cool plasma from the loop apex toward the footpoints. These apparent plane of sky (POS) velocities range between $\mathrm{50 - 100~km~s^{-1}}$ (Figure 11b and 11c). The apparent and Doppler velocities are almost zero near the loop apex and increase towards the footpoints (Figure 11b \& 11c). These velocities are the two orthogonal - LOS ($V_{LOS}$, Dopplergrams) and POS ($V_{POS}$, apparent) components of the downflowing plasma motions with respect to the observer. 

We note that line profiles of bright loop top seen in Ca {\sc{ii}} and H$\beta$ images (Figure 3, 4, 8, 11b) have a very strong central reversal. 
This bright top is formed through the accumulation of evaporated plasma and it can have higher density and optical depth compared to loop legs, which explains why its line profiles show a central reversal \citep{2015ApJ...813..125K,2016ApJ...832..147K,2017ApJ...846....9K}.
We will investigate the density structure of this loop top and formation of central reversal in the follow-up study.

As we mentioned above, the loop footpoints are no longer visible because they have rotated to the far side of the Sun. 
However, the downflows from the loop apex toward the footpoints produce blueshifts above the rightmost footpoint and redshift above the leftmost footpoint (Figure 11a). 
This suggest that the right hand footpoint is located nearer to an observer. 
The measured $B _{LOS}$ is negative everywhere along the loops suggesting that the left footpoint of this arcade has positive polarity and right one has negative.

\subsection{Magnetic flux density}

Since the moving plasma 
must obey  Alfv\'en's theorem
on the scales observed (i. e. be frozen in loops or perhaps in sheets separating
them),
the direction of flow velocity follows the direction of the magnetic loops. The viewing angle, $\gamma$, of the magnetic field and velocity with respect to the observer, (Figure 12), can be determined from the ratio between the LOS and perpendicular components of either the magnetic field or velocity vectors. In contrast to the LOS component, it is 
impossible with given noise levels
to compute the perpendicular component of the magnetic field with the WFA fitting as it is related to the total linear polarization, which depends on the second derivative of Stokes $I$ (see Equation (3)). The measured linear polarization signals are noisy and cannot be used to estimate the field component that is perpendicular to our LOS. The ratio of  $V_{POS}/{V_{LOS}}$ at different parts of the loop system ranges between 1.7 and 4, showing that the dominant component of the downflow motion and hence the magnetic field is the component perpendicular to the LOS (Figures 11b and 12). The average viewing angle of the velocity vectors computed from these ratios is approximately $\gamma \approx$ 60 - 80 degrees (Figure 11b \& 12). This yields a median of total magnetic flux density, $B_{TOT}\sim B_{LOS}/\cos\gamma$ (Figure 12), of around 350 G at the apex of the loop system (region 1 in Figure 8c), and 420 G at mid-heights (region 2 in Figure 8c).

\section{Discussion and Conclusion}

Here we present 2D maps of the magnetic field of a full coronal loop system using chromospheric spectropolarimetry with unprecedented spatiotemporal resolution. 
This is a unique observation with its associated results for the following reasons: (i) the flare took place when the active region was at the West limb, a vantage position that allowed us to see the plasma rise from the chromosphere to coronal heights; (ii) we were able to overcome the challenges posed by the weak signal of coronal lines by using a chromospheric diagnostic at coronal heights; (iii) our observations and magnetic field measurements achieved an unprecedented spatial ($\sim$170 km) and temporal ($\sim$16 seconds) resolution up to a height of 25 Mm into the corona; 
(iv) the high signal allows us to apply the WFA, which is straightforward and does not involve elaborate assumptions or modelling. Yet it provides an accurate estimate of the coronal magnetic field.

Our analysis reveals coronal magnetic field strengths as high as 350 Gauss at heights up to 25 Mm above the solar limb. These values are considerably higher than previous estimates for the coronal field obtained with low-resolution spectropolarimetry and coronal magnetoseismology
\citep{2005LRSP....2....3N,2007Sci...317.1192T,2000ApJ...541L..83L,2004ApJ...613L.177L}. To estimate the effect of spatial and temporal resolution on the measurement of the magnetic field we degraded our data to 1.5$''$ and repeated the measurements. Compared to the original high-resolution data, the LOS magnetic flux density is underestimated by $\sim$70 - 80 \% along different parts of the loops, which partly explains the low values reported in previous works. 

The temporal evolution of the magnetic field maps shows no significant changes over the 30 minute period (between 16:28 and 16:45 UT) (Figures 8, 13 and 14), 
indicating that the loops remain stable even in the aftermath of a large flare, and despite the large mass flux along the loops.
We note that a complementary study using white light data from the HMI instrument on board SDO shows that the density of these post flare loops are around $\mathrm{10^{13}~cm^{-3}}$  
\citep{2018ApJ...867..134J} which is almost 2 order of magnitudes higher than typical coronal loop densities \citep{2009A&A...495..587Y}.

This first high-resolution measurement of the magnetic field strength of solar coronal loops represents a major step forward in understanding coronal magnetism.
The observations and analysis prove that, under certain circumstances, the high-resolution spectropolarimetry of the Ca {\sc{ii}} 8542 {\AA} line gives an accurate measure of the coronal field.   
This constraint is crucial for physical models of coronal active regions, flares and eruptions, and provides a validation of widely-used numerical methods for the extrapolation of photospheric magnetic fields in the corona. Furthermore, the result is important for upcoming new-generation ground-based solar telescopes such as 4-m Daniel K. Inouye Solar Telescope (DKIST) and European Solar Telescope (EST) (first lights in 2020 and 2027, respectively). These telescopes will have advanced chromospheric polarimetric capabilities, which as demonstrated here, can provide powerful diagnostics for the coronal magnetic field.

\begin{acknowledgements}

The research leading to these results has received funding from the S\^er Cymru II Part-funded by the European Regional Development Fund through the Welsh Government. The Swedish 1-m Solar Telescope is operated on the island of La Palma by the Institute for Solar Physics of Stockholm University in the Spanish Observatorio del Roque de los Muchachos of the Instituto de Astrof'sica de Canarias. The Institute for Solar Physics is supported by a grant for research infrastructures of national importance from the Swedish Research Council (registration number 2017-00625). D.K., M.M. and A.R. acknowledge support from STFC under grant number ST/P000304/1. The work of D.K. was supported by Georgian Shota Rustaveli National Science Foundation project FR17\_323. R.O. acknowledges support from MINECO and FEDER funds through projects AYA2014-54485-P and AYA2017-85465-P. 
J.K. acknowledges the project VEGA 2/0004/16. This article was created by the realisation of the project ITMS No.
26220120029, based on the supporting operational Research and development
program financed from the European Regional Development Fund. 
T.V.Z. was supported by the Austrian Science Fund (FWF, project 30695-N27. 
We would like to thank the referee Dr Philip Judge for comments and suggestions that help us improve the presentation of the results in this manuscript.

\end{acknowledgements}

\bibliography{bibtex.bib}
\end{document}